\documentclass{article}
\usepackage[english]{babel}
\usepackage{amsfonts,amssymb, amsmath}

\begin{document}


\title{On novel Hamiltonian description of the nonholonomic Suslov problem }
\author{A.V. Tsiganov\\
\it\small St.Petersburg State University, St.Petersburg, Russia\\
\it\small Beijing Institute of Mathematical Sciences and Applications, Beijing, China\\
\it\small e--mail: andrey.tsiganov@gmail.com}

\date{}
\maketitle

\begin{abstract}
We present some new Poisson bivectors that are invariants by the flow of the nonholonomic Suslov problem. Two rank four invariant Poisson bivectors have globally defined Casimir functions and, therefore, define cubic Poisson brackets on the five dimensional state space with standard symplectic leaves. For the Suslov gyrostat in the potential field we found rank two Poisson bivectors having only two globally defined Casimir functions and, therefore, we say about formal Hamiltonian description in these cases.
 \end{abstract}

\begin{flushright}
Dedicated to the memory of my friend and coathor Alexey Borisov
\end{flushright}

\section{Introduction}
\setcounter{equation}{0}
The Suslov problem \cite{sus02} describes the motion of a rigid body with a fixed point subject
to a nonholonomic constraint  that forces the angular velocity component along a given direction in the body to vanish. 
This nonholonomic system has been the subject of extensive research, see \cite{bbm16,bkm11,drag08, fed09,nar14, nar17,nar24,hu18,jim18,jov01,mac22,zen00} and references therein. 

It allows us to start directly  with a system of autonomous ordinary differential equations   
\begin{equation}\label{m-eq}
\dot{x}_i=X_i(x_1,\ldots,x_5)\,,\qquad i=1,\ldots,5\,,
\end{equation}
where $x=(\gamma_1,\gamma_2,\gamma_3,\omega_1,\omega_2)$ consists of three entries of the
unit vector $\gamma$ and two entries of the angular velocity vector $\omega$,
which are expressed in the special body frame. This frame is firmly attached to the body and its axes are arranged so that the nonholonomic constraint is  
\[\omega_3=0\,,\]
whereas symmetric inertia tensor looks like
\[
I=\left(
    \begin{array}{ccc}
      I_{11} & 0 & I_{13} \\
      0 & I_{22}& I_{23} \\
      I_{13} & I_{23} & I_{33}\\
    \end{array}
  \right)\,,
\]
with  $I_{11}>0$ and  $I_{22}>0$. One more condition must be fulfilled for real bodies \cite{bkm11}.

The vector field $X$ is defined by the  Euler-Poisson equations  
\begin{align}
 &\dot \gamma_1 = -\omega_2\gamma_3, \qquad
\dot \gamma_2 = \omega_1\gamma_3, \qquad
\dot \gamma_3 = \omega_2\gamma_1 -  \omega_1\gamma_2\nonumber
\\
\label{ep-eq}
\\
&I_{11}\dot\omega_1= -( I_{13}\omega_1+ I_{23}\omega_2) \omega_2 ,
\qquad I_{22}\dot\omega_2=( I_{13}\omega_1+ I_{23}\omega_2) \omega_1 \,.
\nonumber
\end{align}
See \cite{bkm11,fed09} for a step-by-step derivation of these Euler-Poisson equations.

According to \cite{fed09} solutions of the equations (\ref{ep-eq}) are meromorphic solutions iff either
\[
I_{13} =0,\qquad  I_{11}=I_{22} +k^2\frac{I_{23}^2}{I_{22}}\,, \]
or
\[
I_{23}  =0,\qquad I_{22}=I_{11}+k^2\frac{I_{13}^2}{I_{11}}\,, \]
where $k$ is a nonzero integer.  In both cases exist three scalar solutions $f_1(x),f_2(x)$ and $f_3(x)$ of the invariance equation  (\ref{scal-eq}) which can be found in \cite{fed09}. 

Below we study generic case without of these restrictions on entries of inertia tensor.  Our aim is to discuss several previously unknown  tensor invariants  $T$ of the flow generated by  $X$ (\ref{ep-eq}) which satisfy to the invariance equation 
\begin{equation}\label{g-inv}
\mathcal L_X\,  T=0\,,
\end{equation} 
Here, $\mathcal L_X T$ is a Lie derivative of tensor field $T$ along vector field $X$ from (\ref{m-eq}) that determines the rate of change of the tensor field $T$ under the state space deformation defined by the flow of $X$. In local coordinates the Lie derivative of the tensor field $T$ of type $(p, q)$ is equal to
\begin{align}
({\mathcal {L}}_{X}T)^{i_{1}\ldots i_{p}}_{j_{1}\ldots j_{q}}=\sum_{k=1}^n X^{k}(\partial _{k}T^{i_{1}\ldots i_{p}}_{j_{1}\ldots j_{q}})
&-\sum_{\ell=1}^n (\partial _{\ell}X^{i_{1}})T^{\ell i_{2}\ldots i_{p}}_{j_{1}\ldots j_{s}}-\ldots - \sum_{\ell=1}^n (\partial _{\ell}X^{i_{p}})T^{i_{1}\ldots i_{p-1}\ell}_{j_{1}\ldots j_{s}}\label{lie-d}\\
&+\sum_{m=1}^n(\partial _{j_{1}}X^{m})T^{i_{1}\ldots i_{p}}_{m j_{2}\ldots j_{q}}+\ldots +\sum_{m=1}^n(\partial _{j_{q}}X^{m})T^{i_{1}\ldots i_{p}}_{j_{1}\ldots j_{q-1}m}
\nonumber
\end{align}
where $\partial _k= {\partial }/{\partial x_k}$ is the partial derivative on the $x_k$ coordinate.

The general theory of tensor invariants is discussed in the following classical books \cite{poi,car}  and in the modern review \cite{koz19}. Different partial solutions of the invariance equation for integrable and non-integrable Hamiltonian systems are discussed in \cite{ts25a, ts25b, ts25c}.

\subsection{Known invariants}
Let us describe known solutions of invariance equation (\ref{g-inv}) for the Suslov problem. In the space of scalar fields $f$ of type $(0,0)$ the invariance equation (\ref{g-inv}) has the form
\begin{equation}\label{scal-eq}
\mathcal L_X f= X^1\frac{\partial f}{\partial x_1}+\cdots+X^5\frac{\partial f}{\partial x_5}=0\,.
\end{equation}
We can solve this equation using the method of undetermined coefficients and obtain the energy and  length of the Poisson vector $\gamma$ as a base in the space of solutions
\begin{equation}\label{f-int}
 f_1=I_{11}\omega_1^2+I_{22}\omega_2^2 \qquad\mbox{and}\qquad f_2=\gamma_1^2+\gamma_2^2+\gamma_3^2\,.
\end{equation}

Divergency of vector field $X$ is equal to
\begin{equation}\label{div-x}
\mbox{div} X=\sum_{k=1}^5 \frac{\partial X_k}{\partial x_k}=\frac{I_{23} }{I_{22}}\omega_1 -\frac{I_{13}}{I_{11}}\omega_2\,.
\end{equation}
Substituting div$X$ into the definition of the Darboux polynomial $D(x)$ as a cofactor
\begin{equation}\label{d-eq}
\mathcal L_X D(x)=c(x)\cdot D(x)\,,\qquad c(x)=\mbox{div} X
\end{equation}
we obtain irreducible Darboux polynomial
\begin{equation}\label{d-pol}
D(x)=I_{13}\omega_1 + I_{23} \omega_2\,.
\end{equation}
If a solution of system (\ref{m-eq}) has a point on the hypersurface $D(x) = 0$,
then the whole solution is contained in this hypersurface, i.e. this invariant hypersurface divides the state space into invariant parts, which makes it easier to study the dynamics of the vector field $X$, see \cite{ll17,mac22}.

Because cofactor $c(x)$ in (\ref{d-eq}) is equal to divergency we have an invariant measure 
\[
\rho(x)=D(x)^{-1}
\]
which is singular on the invariant hypersurface $D(x)=0$,  see discussion in \cite{bkm11,fed09,mac22}. So, there are invariant differential form of type (0,5)
\begin{equation}\label{form}
\Omega=D(x)^{-1}\,dx_1\wedge\cdots\wedge dx_5
\end{equation}
and invariant multivector field of type (5,0)
\begin{equation}\label{mten}
\mathcal E=D(x)\,\frac{\partial}{\partial x_1}\wedge \cdots\wedge \frac{\partial}{\partial x_5}\,.
\end{equation}
Tensor products of invariants $\mathcal E df_{1,2}$ and $\mathcal E df_1df_2$ are also solutions of the invariance equation (\ref{g-inv}).

It is easy to see, that there is one more Darboux polynomial
\[
D_{im}(x)= \sqrt{-I_{11}\,}\omega_1 -\sqrt{I_{22}\,}\omega_2
\]
so that
\[
\mathcal L_XD_{im}(x)=c_{im}(x)D_{im}(x)\,,\qquad c_{im}(x)=\sqrt{-\frac{1}{I_{11}I_{22}}\,} D(x)\,.
\]
The corresponding invariant hypersurface  has no physical meaning.

\section{Rank four invariant Poisson bivectors}
 \setcounter{equation}{0}
The Poisson bivector $P$ is  a contrvariant antisymmetric multivector fields  of valency (2,0) which satisfy to the Jacoby identity
\begin{equation}\label{p-poi}
[\![P,P]\!]=0
\end{equation}
which we represent using the Schouten-Nijenhuis bracket $ [\![.,.]\!]$ on multivector fields.   Any bivector field can be regarded as a skew homomorphism and the rank of this field at a point $x_0$ is the rank of the induced linear mapping  \cite{vai94}. 

Choosing local coordinates $x_1,\ldots,x_n$  any Poisson bivector is given by 
\[P=\sum _{i<j} P^{ij}(x) \frac {\partial }{\partial x_{i} } \frac {\partial }{\partial x_{j}}\,,\]
where $P^{ij}(x)$ is a  skew-symmetric smooth functions. In our partial case we will identify rank of the Poisson matrix $P^{ij}$  with a rank of the Poisson bivector at the point$x_0$.
The Casimir function $C(x)$ of the Poisson bivector $P$ can be defined by the following equation
\[
PdC(x)=0\,.
\] 
The number of the independent Casimir functions of $P$ is related to the rank of $P$ \cite{vai94}.

A Poisson bracket is given in terms of $P$ by
\[ \{f,g\}=Pdf dg=\sum_{i<j} P^{ij}\frac{\partial f}{\partial x_i}\frac{\partial g}{\partial x_j},\]
where $f$ and $g$ are functions on local coordinates $x$. In term of the Poisson bracket the Jacobi condition (\ref{p-poi}) looks like
\[
\{\{f,g\},h\}+\{\{g,h\},f\}+\{\{h,f\},g\}=0\,.
\]
The Poisson brackets between the  Casimir function $C$ and any other function is equal to zero
\[
\{C,f\}=0\,,\qquad \forall f(x)\,.
\]

The Hamilton function $H(x)$  determines the Hamiltonian vector field $X$  
\begin{equation}\label{ham-eq}
X=PdH\,,\qquad\mbox{or}\qquad X(g)=\{H,g\}\,.
\end{equation}
The Hamiltonian vector fields  $X$  generate an integrable generalized distribution and the leaves of this foliation are symplectic. Usually the flow of the Hamiltonian vector field preserves the Poisson structure, it fixes each leaf and the Hamiltonian itself is a first integral. 

So, we can try to compute Poisson bivector $P$ as a tensor invariant of a given vector field $X$, i.e. as a solution of invariance equation  (\ref{g-inv})
\[
\mathcal L_X P=0\,.
\]
The scalar invariant $H$ (\ref{ham-eq}) is not fixed in this approach. It is computed later using the invariance property.

\subsection{Cubic Poisson brackets}
In the space of arbitrary multivector fields $T$ of valency $(2,0)$ the invariance equation looks like
\begin{equation}\label{t-eq}
(\mathcal L_X T)^{ij}=\sum_{k=1}^5\left(X^k\frac{\partial T^{ij}}{\partial x^k}-T^{kj}\frac{\partial X^i}{\partial x^k}-T^{ik}\frac{\partial X^j }{\partial x^k} \right)=0\,.
\end{equation}
We solve this equation using the method of undetermined coefficients when we suppose that the entries of the multivector field $T(x)$ are inhomogeneous  polynomials on variables $x_1,\ldots,x_5$ of degree $N$.

At $N=1$ we have the following known rank two invariant bivector
\begin{equation}\label{P1}
P_1=\left(\begin{array}{ccccc}
0& \gamma_3& -\gamma_2& 0& 0\\-\gamma_3& 0& \gamma_1& 0& 0\\ \gamma_2& -\gamma_1& 0& 0& 0\\0& 0& 0& 0& 0\\0& 0& 0& 0& 0
\end{array}\right)\,.
\end{equation}
We use the notation $P_1$ instead of $T_1$ because this bivector satisfies the Jacobi condition (\ref{p-poi}), 
 that allows as introduce the following linear Poisson bracket
\[
\{\gamma_i,\gamma_j\}_1=\epsilon_{ijk}\gamma_k\,.
\]
Here $\epsilon$ is the totally antisymmetric Levi-Civita tensor. 

At $N=2$ we have no solutions of the invariance equation (\ref{t-eq}).

At $N=3$ we have the following  rank four invariant bivector
\begin{equation}\label{T3}
T_3=(c_1f_1+c_2f_2+c_3)P_1+c_4P_2+c_5P_3\,,\qquad c_k\in\mathbb R\,.
\end{equation}
Here $P_1,P_2$ and $P_3$ are the Poisson bivectors which satisfy to the Jacobi condition (\ref{p-poi}). They are incompatible with each other and, therefore, $T_3$ is the symmetric skew multivector field of valence (2,0) which does not satisfy the Jacobi condition 
\[[\![T_3,T_3]\!]\neq 0\,.\]
Nevertheless it is geometric structure which is invariant with respect to the Suslov model flow.

The entries of the Poisson bivectors $P_2$ and $P_3$ are equal to:
\begin{align}
P_2^{12}&=-\gamma_3 (\omega_1  \gamma_1 + \omega_2   \gamma_2)\,,\qquad P_2^{13}=  \gamma_1 \gamma_2 \omega_1 -(\gamma_1^2 + \gamma_3^2) \omega_2  \,,\qquad
P_2^{23}=(\gamma_2^2 + \gamma_3^2) \omega_1  - \gamma_1 \gamma_2 \omega_2  \,,\nonumber\\
P_2^{14}&=\phantom{-}I_{11}^{-1} \gamma_1 \omega_2   D(x)\,,\qquad
P_2^{15}=-I_{22}^{-1}\gamma_1 \omega_1  D(x)\,,\qquad P_2^{24}=\phantom{-}I_{11}^{-1} \gamma_2 \omega_2   D(x)\,,\nonumber\\
P_2^{25}&=-I_{22}^{-1} \gamma_2 \omega_1  D(x)\,,\qquad P_2^{34}=\phantom{-}I_{11}^{-1} \gamma_3 \omega_2   D(x)\,,\qquad
P_2^{35}=-I_{22}^{-1} \gamma_3 \omega_1  D(x)\,,\nonumber\\
P_2^{45}&=0\,,\label{P2}
\end{align}
and
\begin{align}
P_3^{12}&=0\,,\qquad P_3^{13}=0\,,\qquad P_3^{23}=0\,,\qquad P_3^{14}=-\omega_1\omega_2\gamma_3\,,\qquad P_3^{15}=-\omega_2^2\gamma_3\,,\nonumber\\
P_3^{24}&=\omega_1^2\gamma_3\,,\qquad P_3^{2,5}=\omega_1\omega_2\gamma_3\,,\qquad P_3^{34}=\omega_1(\omega_2\gamma_1 -\omega_1\gamma_2)\,,\qquad
P_3^{35}=\omega_2(\omega_2\gamma_1 -\omega_1\gamma_2)\,,\nonumber\\
P_3^{45}&=-\frac{(I_{11}\omega_1^2 +I_{22}\omega_2^2)D(x)}{I_{11}I_{22}}\label{P3}\,.
\end{align}
Here $D(x)$ is the Darboux polynomial (\ref{d-pol}). 

Substituting generic solution  $T_3$ (\ref{T3}) of the equation (\ref{t-eq}) into the Jacobi condition (\ref{p-poi})
\[
[\![T_3,T_3]\!]=0
\]
and solving the resulting equations on coefficients $c_1,\ldots,c_5$ we obtain the Poisson bivectors at
\begin{enumerate}
  \item $c_1 = 0,\qquad c_2 = c_2,\qquad c_3 = c_3,\qquad c_4 = 0,\qquad c_5 = c_5$;
   \item $c_1 = 0, \qquad c_2 = 0,\qquad  c_3 = 0,\qquad  c_4 = c_4,\qquad  c_5 = c_5$;
  \item $c_1 = c_1,\qquad c_2 = c_2,\qquad  c_3 = c_3,\qquad c_4 = 0,\qquad  c_5 = 0$;
   \item $c_1 = c_1,\qquad  c_2 = 0,\qquad  c_3 = 0,\qquad  c_4 = 2c_5, \qquad c_5 = c_5$.
\end{enumerate}
In the first and last cases, the obtained Poisson bivectors  have  rank four. In the second and third cases, the corresponing Poisson bivectors have  rank  two.

In the first case the  rank four invariant  Poisson bivector is equal to
\begin{equation}\label{Pa}
P_a=(c_2f_2+c_3)P_1+c_5P_3\,.
\end{equation}
A tensor operation between invariant tensor fields results in either a tensor invariant or zero. Therefore, we easy compute the following relations
\[
P_adf_1=2c_5f_1X\qquad\mbox{and}\qquad P_adf_2=0\,.
\]
So, function $f_2=|\gamma|^2$ is the globally defined Casimir function of $P_a$ and the original vector field $X$ has the standard  Hamiltonian form
\[
X=P_adH_a
\]
where Hamilton function
\[
H_a=\frac{1}{2c_5}\ln f_1\,,
\]
is the logarithm of energy up to a constant. If we set $c_5=1/4$, then we get  $H_a=\ln f_1^2$. This is a globally defined function both for positive and negative values of energy.

In a similar way for the second  rank four Poisson bivector
\begin{equation}\label{Pb}
P_b=c_1f_1P_1+2c_5P_2+c_5P_3
\end{equation}
we have
\[
P_bdf_1=2c_5f_1X\,,\qquad P_bdf_2=4c_5f_2X\,.
\]
It allows us to find the globally defined Casimir function
\[
C_b=\ln f_1^2+\ln f_2\,,\qquad P_bdC_b=0\,,
\]
since $f_1^2>0$ and $f_2>0$, and two equivalent  Hamiltonian description of the original vector field
\[
X=P_bdH^{(1,2)}_b\,,\qquad H_b^{(1)}=\frac{\ln f_1}{2c_5}\,,\quad H_b^{(2)}=-\frac{\ln f_2}{4c_5}\,.
\]

\section{Special case of the Euler-Poisson equations}
 \setcounter{equation}{0}
The Euler equation  describes the evolution of  the angular velocity vector $\omega=(\omega_1,\omega_2,\omega_3)$ 
\[
 I \dot{\omega}+\omega\times I\omega=\mu r\times \gamma\,,
\]
Here  $r = (r_1, r_2, r_3)$ and $\gamma=(\gamma_1,\gamma_2,\gamma_3)$  are radius vector of the center of mass and  unit Poisson vector in the system of principal axes, $I = \mbox{diag}(I_1, I_2, I_3)$ is a tensor of inertia, $\mu = mg$ is a  weight of the body and $\times$ means a vector product, see all the details in textbook \cite{bm05}.

The  Poisson equation describes evolution of the Poisson vector $\gamma$
\[
\dot{\gamma}=\gamma\times \omega\,,
\]
For a free rigid body $\mu=0$ and the Euler equation becomes independent of the Poisson equation. The Euler-Poisson equations for the Suslov problem (\ref{ep-eq}) are another example of equations that can be separated  into two parts.

The third example is the Euler-Poisson equations
\begin{equation}\label{ab-eq}
 I \dot{\omega}=\omega\times A\omega \qquad \mbox{and}\qquad \dot{\gamma}=\gamma\times B\omega\,,
\end{equation}
where $A,B$ are two constant $3\times 3$ matrices.  The system of equations (\ref{ab-eq}) describes the partial motion of a rigid body with a fluid-filled cavity, see \S 3 of Chapter 3 in \cite{bm05}. This is the case when the intensity of the fluid vortex in the cavity is small compared to the kinetic moment, or vice versa. These equations  also arise in other  physical applications, see \cite{bt96,bt97,omni} and references within.

The first equation in (\ref{ab-eq}) is independent of the second equation which has zero divergency 
\[\sum_{i=1}^3 \frac{\partial \left(\gamma\times BM\right)_i}{\partial \gamma_i}=0\,,\]
so the phase flow of $X$ preserves the corresponding trivector $\mathcal E_\gamma$, vector field $Y$ and scalar invariant $F_1$ defined by
\[\mathcal E_\gamma=\dfrac{\partial}{\partial \gamma_1}\wedge\dfrac{\partial}{\partial \gamma_2}\wedge \dfrac{\partial}{\partial \gamma_3}\,,\qquad 
Y=\sum_{i=1}^3 \gamma_i\dfrac{\partial}{\partial \gamma_i}\,,\qquad F_1=(\gamma,\gamma)\,.
\]
If  $A$ is a symmetric matrix,  $A=A^T$, the  divergency of the vector field  $X$ (\ref{ab-eq}) is equal to zero
\[
\mbox{div}\,X=
\sum_{i=1}^3 \frac{\partial \left(M\times AM\right)_{i}}{\partial M_i}+
\sum_{i=1}^3 \frac{\partial \left(\gamma\times BM\right)_i}{\partial \gamma_i}\,,
\]
and, therefore, the phase flow of $X$ preserves the volume form 
\begin{equation}\label{inv-vol}
\Omega=d\gamma_1\wedge d\gamma_2\wedge d\gamma_3 \wedge dM_1\wedge dM_2 \wedge dM_3\,.
\end{equation}
 In this case, the first equation in (\ref{ab-eq})  coincides with the  Euler equation describing free rigid body motion
\[\dot{M}=M\times \omega\,,\qquad \mbox{where}\qquad\omega=AM\,,\]
having two scalar invariants
\[
F_2=(M,M)\,,\qquad F_3=\frac{1}{2}(M,AM)\,.
\]
Substituting the well-known solutions $M(t)$ of the Euler equation into the second equation in (\ref{ab-eq}) we non-autonomous Carath\'{e}odory differential equations of the form
\[
\dot{\gamma}=\gamma\times BM(t)\,.
\]
In the mathematical theory of control such systems are known as linear switched systems of differential equations.

According to the Euler–Jacobi–Lie theorem \cite{koz13} if the system of ordinary differential equations  
\[\dot{x}=X(x_1,\ldots,x_n)
\]
on the $n$-dimensional smooth phase space with coordinates $x=(x_1,\ldots,x_n)$ possesses
\begin{enumerate}
  \item $k$ functionally independent first integrals $F_1,\ldots,F_k$; 
       \item $m=n - k - 2$ independent invariant symmetry fields $Y_{k+1},\ldots, Y_{n-2}$
         which together with the field $Y_{n-1} = X$ generate a nilpotent Lie algebra of vector fields;
  \item an invariant volume $n$-form $\Omega$;
        \[\mathcal L_X\Omega=0\,.\]
\end{enumerate}
so that
\[\mathcal L_{Y_i} F_j = 0,\qquad  \mathcal L_{Y_i}\Omega = 0\,,\]
then it is integrated by quadratures.   

In our case, we have $k=3$ independent first integrals $F_1$, $F_2$ and $F_3$,  $m=6-3-2=1$ vector field $Y$ commuting with $X$ and invariant volume form $\Omega$ (\ref{inv-vol}). However, the vector field $X$ (\ref{ab-eq}) does not satisfy the conditions of the Euler–Jacobi–Lie theorem since
\[
\mathcal L_YF_1=2F_1\,.
\]
So, in the generic case, we have a non-integrable system.
\subsection{Formal and informal Hamiltonization}
We have the following tensor invariants of  $X$ (\ref{ab-eq}) : 
\begin{itemize}
  \item three scalar invariants 
  \begin{equation}\label{inv-sc}
  F_1=(\gamma,\gamma)\,,\qquad F_2=(M,M)\,,\qquad F_3=\frac{1}{2}(M,AM)\,;  
  \end{equation}
  \item three invariant  multivectors 
  \begin{equation}\label{inv-mv}
  Y=\sum_{i=1}^3 \gamma_i\dfrac{\partial}{\partial \gamma_i}\,,\qquad 
  \mathcal E_\gamma=\dfrac{\partial}{\partial \gamma_1}\wedge\dfrac{\partial}{\partial \gamma_2}\wedge \dfrac{\partial}{\partial \gamma_3}\,,
  \end{equation}
  and 
  \[
  \mathcal E=\dfrac{\partial}{\partial \gamma_1}\wedge\dfrac{\partial}{\partial \gamma_2}\wedge \dfrac{\partial}{\partial \gamma_3}\wedge \dfrac{\partial}{\partial M_1}\wedge\dfrac{\partial}{\partial M_2}\wedge \dfrac{\partial}{\partial M_3}\,.\]
\end{itemize}
 The Lie derivatives $\mathcal L_X$ of all these invariants are equal to zero.  We will use these invariants to construct Hamilton representation of $X$
 \[ X=PdH\]
 where $H$ is a Hamilton function and $P$ is a Poisson bivector.
 
Another main ingredient is the semi invariant or Darboux invariant $Z$ of the homogeneous vector field $X$
\[
\mathcal L_XZ=X\,,
\]
where vector field
\begin{equation}\label{eul-v}
Z= \sum_{i=1}^3 \gamma_i\dfrac{\partial}{\partial \gamma_i}+ \sum_{i=1}^3 M_i\dfrac{\partial}{\partial M_i}
\end{equation}
is the well-known Euler vector field whose flow  is the linear multiplication action. Since the first integrals $F_k$ are homogeneous functions we have
\[
\mathcal L_ZF_k=2F_k\,,\qquad k=1,2,3,
\]
according to the Euler theorem on homogeneous functions. Moreover, the invariant rank-two bivector
\[
P_e=X\wedge Z\,, \qquad \mathcal L_XP=0\,,
\]
satisfies to the Jacobi condition
\[
[\![P_e,P_e]\!]=0\,,
\]
according to the same Euler theorem. 

The invariant bivector $P$ does not have four independent Casimir functions since we have only three single-valued scalar invariants $F_1,F_2$ and $F_3$,  see discussion in \cite{bizkoz}. So, we can say that this rank-two bivector yields formal Hamiltonization of $X$.

It is simple to confirm that the rank four bivector
\begin{equation}\label{ab-P}
P=(c_1F_1+c_2F_2+c_3F_3)P_\gamma+c_4P_c+eP_e\,,\qquad c_k,e\in\mathbb R\,,
\end{equation}
where
\[
P_\gamma=\mathcal E_\gamma dF_1\,,\qquad P_c=X\wedge Y \qquad\mbox{and}\qquad P_e=X\wedge Z\,. 
\]
is the invariant bivector
\[\mathcal L_XP=0\,,\]
which satisfies the Jacobi condition
\[
[\![P,P]\!]=0\,,
\]
and the following equations hold
\begin{equation}\label{3-eq}
PdF_1=2(c_4+e)F_1X\,,\qquad PdF_2=2eF_2X\,,\qquad PdF_3=2eF_3X\,.
\end{equation}
\
From (\ref{3-eq}) follows that we can define two independent Casimir functions 
\[
C_1=\left(1+\frac{c_4}{e}\right)\ln F_2-\ln F_1\,,\qquad
C_2=\left(1+\frac{c_4}{e}\right)\ln F_3-\ln F_1\,.
\]
This means that $P_D C_1,2 = 0$, and the vector field $X$ (\ref{ab-eq}) can be rewritten in Hamiltonian form
\[
X=PdH\,,\qquad H=\frac{1}{e}\ln F_2\,.
\]
Here $P$ is a rank four Poisson bivector with  two well-defined  Casimir functions $C_{1,2}$ and $F_2$ is a mechanical energy. We can therefore talk about the informal Hamiltonisation of the vector field $X$.

If $A$ is an arbitrary matrix $A\neq A^T$, the divergence of the vector field $X$ (\ref{ab-eq}) does not equal zero.
\[
\mbox{div}\,X=(a_{2,3} - a_{3, 2})M_1 + (a_{3,1}-a_{1, 3})M_2 + (a_{1, 2} - a_{2, })M_3 
\]
and, therefore, the phase flow of $X$ does not preserve the volume form $\Omega$ (\ref{inv-vol}) and function
\[F_3=\frac{1}{2}(M,AM)\,.\]
The rank four invariant Poisson bivector (\ref{ab-P}) now depends on four parameters $c_k$
\begin{equation}\label{ab-P2}
P=(c_1F_1+c_2F_2)P_\gamma+c_3P_c+eP_e\,,\qquad c_k,e\in\mathbb R\,.
\end{equation}
But we cannot now construct two independent single-valued Casimir functions $C_{1,2}$ and the Hamiltonian $H$ from only two scalar invariants $F_{1,2}$ of the phase flow. So, in this case, we can only say about the formal Hamiltonization of the vector field $X$.

\section{Conclusion}

Our aim is to consider a system of 
ordinary differential equations
\[
\dot{\mathrm{x} }_k =X_k(\mathrm{x}_1,\ldots,\mathrm{x}_N)\,.
\]
having the same structure as the Euler-Poisson equations for the free top, for the Suslov problem and for some other known systems of equations from the textbook \cite{bm05}. For all these systems we can divide $N$ variables $\mathrm{x}=(\mathrm{x}_1,\ldots,\mathrm{x}_N)$ on two parts
\[
x=(x_1,\ldots,x_n)\qquad \mbox{and}\qquad y=(y_1,\ldots,y_m)\,,\qquad n+m=N
\]
and consider a system of equations  with quadratic right-hand sides
\begin{equation}\label{xy-eq}
\dot{x}_k =\sum_{i,j=1}^n a_k^{ij} x_ix_j\qquad\mbox{and}\qquad 
\dot{y}_\ell =\sum_{i=1}^n\sum_{j=1}^m b_\ell^{ij} x_iy_j\,,\qquad a_k^{ij},b_\ell^{ij}\in\mathbb R\,.
\end{equation}
We also suppose that first subsystem of equations has $n-1$ independent  first integrals $f_1,\ldots,f_{n-1}$ depending on $x_k$, whereas second subsystem of equations has $m-2$ independent first integrals $g_1,\ldots,g_{n-2}$ depending on $y_\ell$.

The intersection of the invariant levels of the first integrals $f_1,\ldots,f_n$ determine solutions $x_k(t)$. When we  substitute these solutions  into the second subsystem, we obtain non-autonomous system of $m$ equations
\[
\dot{y}_\ell =\sum_{i=1}^n\sum_{j=1}^m b_\ell^{ij} x_i(t)y_j\,,\qquad \ell=1,\ldots,m,
\]
which has $m-2$ first integrals $g_1,\ldots,g_{m-1}$. It allows us to study solutions of this counterpart of the Poisson equations in  the framework of the Carath\'{e}odory theory. 

We want to find an invariant Hamiltonian representation of differential equations (\ref{xy-eq}). In this note we consider two partial cases of equations (\ref{xy-eq}) with $N=5$ and $N=6$. A general case will be considered in subsequent publication.

 \section*{Funding}
The study was carried out with the financial support 
of the Ministry of Science and Higher Education of the Russian Federation in the framework of a scientific project under agreement No 075-15-2025-013 by St. Petersburg State University as part of the national project “Science and Universities” in 2025.


\begin{thebibliography}{99}
\bibitem{bizkoz}
Bizyaev I.A., Kozlov V.V.,
\newblock{Homogeneous systems with quadratic integrals, Lie-Poisson quasibrackets, and Kovalevskaya's method},
Sbornik: Mathematics, 206:12 (2015), 1682-1706.

\bibitem{bbm16}
Bizyaev, I.A.,  Borisov, A.V., Mamaev, I.S., The Hojman construction and Hamiltonization of nonholonomic systems, \newblock{\em SIGMA}, 12 (2016), 012, 19 pp. 

\bibitem{bm05} 
Borisov A.V.,  Mamaev I.S., {\em Dynamics of a rigid body}, Regular and Chaotic
Dynamics Publ., Moscow-Izhevsk (2001), 384 pp.

\bibitem{bt96}
Borisov, A. V., Tsygvintsev, A.V., Kowalewski Exponents and Integrable Systems of Classic
Dynamics: 1, 2, {\em Regul. Chaotic Dyn.}, 1:1 (1996), 15-37.

\bibitem{bt97}
Borisov, A. V., Tsygvintsev, A.V., Kovalevskaya’s Method in Rigid Body Dynamics, 
{\em J. Appl. Math. Mech.}, 61:1 (1997), 27-32.

\bibitem{bkm11}
  Borisov, A. V., Kilin, A.A., and Mamaev, I. S., Hamiltonicity and Integrability of the Suslov Problem,
\newblock{\em Regul. Chaotic Dyn.},  16:1-2 (2011),  104-116.

\bibitem{omni}
 Borisov A.V.,  Kilin A.A., Mamaev I.S., Dynamics and control of an
omniwheel vehicle, {\em Regul. Chaotic Dyn.}, 20:2 (2015), 153-172.

\bibitem{car}
Cartan, E., \newblock{\em Le\c{c}ons sur les invariants int\'{e}graux}, Hermann, Paris, 1922, 210 pp.


  
\bibitem{drag08}
 Dragović, V., Gajić, B. Hirota-Kimura type discretization of the classical nonholonomic Suslov problem, \newblock{\em  Regul. Chaot. Dyn.}, 13 (2008), 250-256.
 
  \bibitem{fed09} 
 Fedorov, Yu.N., Maciejewski, A. J., and Przybylska, M., Suslov Problem: Integrability, Meromorphic
and Hypergeometric Solutions, \newblock{\em  Nonlinearity},  22 (2009), 2231-2259.
 
 \bibitem{nar14}
 García-Naranjo, L.C.,  Maciejewski, A. J.,  Marrero, J.C., Przybylska, M.,
The inhomogeneous Suslov problem,
\newblock{\em Phys.s Lett. A},  378 (2014), 32-33.
 
 \bibitem{nar17}
García-Naranj, L.C., F. Jiménez, F., The geometric discretisation of the Suslov problem: A case study of consistency for nonholonomic integrators,
\newblock{\em  Discr. Cont. Dyn. Systems},  37:8 (2017), 4249-4275.
 
 \bibitem{nar24}
 García-Naranjo, L.C., Marrero, J.C., de Diego, M.,  Valdés, P.,  Almost-Poisson Brackets for Nonholonomic Systems with Gyroscopic Terms and Hamiltonisation,
  \newblock{\em  J Nonlinear Sci.}, 34 (2024), 110.
 
 
 \bibitem{hu18}
Hu, S., Santoprete, M., Suslov Problem with the Clebsch-Tisserand Potential,
\newblock{\em  Regul. Chaot. Dyn.}, 23 (2018), 193-211.
 
 \bibitem{jim18}
 Jiménez, F., Scheurle, J.,  On some aspects of the discretization of the Suslov problem, 
\newblock{\em J. Geom.  Mech.}, 10:1, (2018), 43-68.
 
\bibitem{jov01}
Jovanovi\'{c}, B. Geometry and Integrability of Euler-Poincar\'{e}-Suslov Equations,  \newblock{\em Nonlinearity},
14 (2001), 1555-1657.

\bibitem{koz13}
Kozlov V. V., {The Euler-Jacobi-Lie integrability theorem}, 
\newblock{ Regul. Chaotic Dyn.}, {18}:4, (2013), 329-343.
  
\bibitem{koz19}
Kozlov, V. V.,
{Tensor invariants and integration of differential equations},
\newblock{\em  Russ. Math. Surv.}, 74:1 (2019), 111-140.

 \bibitem{lie}
Lie, S., \newblock{\em Theorie der transformationsgruppen}, Zweiter Abschnitt, unter mitwirkung von Prof.
Dr. Friedrich Engel, Teubner, Leipzig, 1890.

 \bibitem{ll17}
Llibre, J., Valls, C. Darboux polynomials, balances and Painlev\'{e} property, \newblock{\em  Regul. Chaot. Dyn.}, 22 (2017), 543-550.

\bibitem{mac22}
 Maciejewski, A.J.,  Przybylska, M., Gyrostatic Suslov Problem, \newblock{\em Rus. J. Nonlin. Dyn.}, 18:4 (2022), 609-627.

\bibitem{poi}
Poincar\'{e}, H.,
\newblock{\em  Les M\'{e}thodes Nouvelles de la M\'{e}canique C\'{e}leste}, Tom III, Dover Pub. Inc., N.Y., (1892).

\bibitem{sus02} Suslov, G.  {\em Theoretical Mechanics}, Vol. 2, 1902, Kiev (in Russian).

\bibitem{ts25a}
Tsiganov, A.V.,
On tensor invariants for integrable cases of Euler, Lagrange and Kovalevskaya rigid body motion, 	
\newblock{\em Izv. Ross. Akademii Nauk. Ser. Matem.},  89:2 (2025), 161-188.


\bibitem{ts25b}
Tsiganov, A.V., On tensor invariants of the Clebsch system,
{\em Reg. Chaot. Dyn.}, 30:4 (2025), 742–764.

\bibitem{ts25c}
Tsiganov, A.V., Homogeneous potentials, Lagrange's identity and Poisson geometry, Preprint: 	arXiv:2511.19903, (2025).

\bibitem{vai94}
Vaisman, I., {\em Lectures on the Geometry of Poisson Manifolds}, Birkh\"{u}auser, Basel, 1994.


\bibitem{zen00}
Zenkov, D.V.,  Bloch, A.M.,  Dynamics of the $n$-dimensional Suslov problem, \newblock{\em  J. Geom. Phys.}, 34 (2000), 121-136.
\end{thebibliography}
\end{document}